\documentclass{article}
\usepackage{graphicx}
\begin{document}

\title{Variations of Gauss-Codazzi-Ricci Equations in Kaluza-Klein Reduction (String Theory) and Cauchy Problem (General Relativity)}
\author{Pei Wang \thanks{peiwang@nwu.edu.cn} \\
Institute of Modern Physics, Northwest University, Xian 710069,
China}

\date{}
\maketitle

\begin{abstract}
We find a kind of variations of Gauss-Codazzi-Ricci equations
suitable for Kaluza-Klein reduction and Cauchy problem. Especially
the counterpart of extrinsic curvature tensor has antisymmetric
part as well as symmetric one. If the dependence of metric tensor
on reduced dimensions is negligible it becomes a pure
antisymmetric tensor. {PACS:03.70;11.15} {\it Keywords:} Gauss
Codazzi Ricci equation, Kaluza Klein reduction

\end{abstract}

\section{Introduction}
\subsection{Introductory Remarks}
As is well-known Gauss-Codazzi-Ricci equations are very important
instruments for describing a submanifold in a Riemann space. By
nature they appear in the Cauchy problem of general relativity
\cite{Hawking}, \cite{Wald}.However the celebrated Kaluza-Klein
dimensional reduction in string/M theory involves also a
submanifold in a higher dimensional space-time. Therefore it is
interesting to find the variations of Gauss-Codazzi-Ricci
equations suitable for dimensional reduction. It would be best to
look for a formulation which is convenient for both purposes:
Cauchy problem in general relativity and Kaluza-Klein reduction in
string/M theory.\\There are authors who proposed to use the lapse
function\cite{Arnowitt} and shift vector function\cite{Wheeler}
for characterizing the Cauchy problem as well as Hamiltonian
formulation in general relativity. As shown in
refs.\cite{Hawking,Wald} n-dimensional (pseudo) Riemann space can
be foliated by a family of n-1 dimensional Cauchy surfaces. On
these hypersurfaces there exists "normal vector field" which is
connected with the extrinsic curvature. But because of the shift
vector the covariant form $n_A(A=1,\cdots\,n)$ of normal vector is
not in the same direction with contravariant vector $n^A$. They
like the pair of axes of a oblique coordinates, $n_A$ is
orthogonal to the initial surface($x^n=0$) but $n^A$ is not (see
next subsection for the detail).\\Now we suggest a similar
construction with lapse-like function related to the dilaton
fields in string theory and shift-like vector function connected
with Kaluza-Klein gauge field. Of course in general relativity we
can use them to determine the Hamiltonian structure as well. In
this construction the role of $n_A$ and $n^A$ are exchanged. The
vector orthogonal to the initial surface is $n^A$ instead of
$n_A$.
\\Significantly, the new construction leads to variations of
Gauss-Codazzi-Ricci equations. Especially the counterpart of the
extrinsic curvature has antisymmetric part and symmetric one. If
the dependence of metric tensor on reduced dimensions can be
neglected it becomes a pure antisymmetric tensor. This is what we
will depict in this paper. \\In second part of this section we
would like to give a brief review about lapse function and shift
vector in order to compare them with new construction. We restrict
ourselves from sec.2 to 4 only one reduced dimension included.
Section 2 is devoted for Gauss and Weingarten formulae while Gauss
and Codazzi equations are presented in sec.3 and sec.4
respectively. In sec.5 we discuss this construction presented in
general relativity. Sec.6 collects the main interest in the
dimensional reduction including reduced dimensions more than one.
Finally a short discussion about future work will appear in sec.7.
Besides we list the relevant Christoffel symbols in an Appendix.

\subsection{Lapse Function and Shift Vector
Review\cite{Wald},\cite{Stephani}}

Einstein's theory asserts that space-time structure and
gravitation are described by a pseudo Riemann space and a metric
tensor $g_{AB}$. It has been proved that n-dimensional (pseudo)
Riemann space can be foliated by a family of n-1 dimensional
Cauchy surfaces. On these hypersurfaces there exists a normal
vector field $n_A$ satisfying the normalized condition
\begin{equation}n_An^A=\epsilon=\pm 1,\end{equation}
+1 and -1 corresponding to the space-like and time-like vector
respectively. When the lapse function N and shift function
$N_\alpha(\alpha=1\cdots\,m)$ (in sec.2-4 m=n-1) are introduced,
using a coordinate frame, vector $n_A$ can be denoted by its
components
\begin{equation}n_A=(\underbrace{0,\cdots,0}_{n-1},\epsilon\,N),\end{equation}
\begin{equation}n^A=(-N^\alpha/N,1/N).\end{equation}
 Let $h_{AB}$ represent the metric on each
hypersurfaces induced by $g_{AB}$. Then we have
\begin{equation}g_{AB}=h_{AB}+\epsilon\,n_An_B,\end{equation}
\begin{equation}h_{AB}n^B=h^{AB}n_B=0.\end{equation}
 Explicitly we can write down inverse metric
\begin{equation}g^{AB}=\left(\begin{array}{cc}
h^{\alpha\beta}+{\epsilon\,N^\alpha\,N^\beta}/N^2 &
-{\epsilon\,N^\alpha}/N^2 \\
-{\epsilon\,N^\beta}/N^2 & \epsilon/N^2\end{array}\right),\end{equation}\\
or
\begin{equation}
h^{AB}=\left(\begin{array}{cc}h^{\alpha\beta}& 0 \\
0&0\end{array}\right)\end{equation}\\
and others. It is worthy to note that we have $h^n_A=0$ which is
the most remarkable distinction with later construction.\\
The extrinsic curvature tensor is defined by
\begin{equation}K_{AB}=h^\alpha_Ah^\beta_B\,K_{\alpha\beta},\end{equation}
\begin{equation}K_{\alpha\beta}=\epsilon\,N\Gamma_{\alpha\beta}^n\\
=\frac{\epsilon}{2N^2}[(\mathbf{\nabla
}_\alpha\,N_\beta+\mathbf{\nabla}_\beta\,N_\alpha)-\partial_nh_{\alpha\beta}],\end{equation}
\begin{equation}\mathbf{\nabla}_\alpha\,N_\beta\equiv\partial_\alpha-\mathbf{\Gamma}_{\alpha\beta}^{\gamma}
\,N_{\gamma}\end{equation} which is a symmetric tensor connected
with the second fundamental form of the hypersurface,
$\Gamma_{AB}^C$ and $\mathbf{\Gamma}_{\alpha\beta}^{\gamma}$ are
n-dimensional and m(=n-1)-dimensional Christoffel symbol
respectively.\\From about definitions one can derive the
Gauss-Codazzi equations and solve the Einstein equation in vacuum
through Hamiltonian formulation either. Please find the detail in
refs\cite{Wald},\cite{Stephani}.
\section{Gauss Formula and Weingarten Formula}
The standard Kaluza Klein reduction from n-dimensional space-time
to n-1 dimensional subspace is shown in the following formula
\begin{equation}ds_n^2=e^{2a\phi}ds_{n-1}^2+e^{2b\phi}(dx^n+\mathcal{A}_\alpha\,dx^\alpha)^2,\end{equation}
\begin{equation}ds_{n-1}^2=h_{\alpha\beta}dx^\alpha\,dx^\beta\end{equation}
in which $\phi$ is the dilaton field and $\mathcal{A}_\alpha $is a
gauge field. Apart from a conformal factor $e^{2a\phi}$ the metric
is in the form
\begin{equation}ds^2=g_{AB}dx^A\,dx^B=h_{\alpha\beta}dx^\alpha\,dx^\beta+\frac{\epsilon}{N^2}(dx^n+N_\alpha\,dx^\alpha)^2,\end{equation}
where \begin{equation}N_\alpha=\mathcal{A}_\alpha \qquad
N^{-2}=e^{2(b-a)\phi} \equiv\, e^{2\mathbf{b}\phi}.\end{equation}
In eq.(13) we have added a factor $\epsilon$ so that it can fit
Cauchy problem as well. We also introduce the normal vector $n^A$
such that
\begin{equation}g_{AB}=h_{AB}+\epsilon\,n_A\,n_B \quad
n_An^A=\epsilon \quad h_{AB}n^B=h^{AB}n_B=0,\end{equation} and
denote it in its components form
\begin{equation}
n^A=(0,\cdots,\epsilon\,N),\qquad
n_A=(N_\alpha/N,1/N).
\end{equation}
Similar to the lapse-shift
case we can write down inverse metric
\begin{equation}
g^{AB}=\left (
\begin{array}{cc}
h^{\alpha\beta}& -N^\alpha\\
-N^\beta & N_\gamma\,N^\gamma+\epsilon \, N^2
\end{array}\right )
= h^{AB}+\left (\begin{array}{cc}0 & 0\\
0 & \epsilon \, N^2\end{array}\right )
\end{equation}
and
\begin{equation}h_{AB}=\left(\begin{array}{cc}h_{\alpha\beta}&0\\0&0\end{array}\right).\end{equation}
It is easy to find that
\begin{equation}h_{n\alpha}=h_{\alpha\,n}=h_n^\alpha=0,\qquad
h_\alpha^\beta=\delta_\alpha^\beta.\end{equation} We may imitate
ref.\cite{Wald} to introduce the "time flow" vector field
$$t_A=(0,\cdots 0,1),\qquad
t^A=(-N^\alpha,N_\gamma\,N^\gamma+\epsilon\,N^2),$$ so that
$$n_At^A=n^At_A=\epsilon\,N,$$ and $$h_{AB}t^B=-N_\alpha.$$ But we
have no chance to use it latter.\\
From eqs.(15)-(18) we know that
\begin{equation}h_{\alpha\beta}=h_\alpha^A\,h_\beta^Bg_{AB}.\end{equation}
Differentiate it we get
\begin{equation}\partial_\gamma\,h_{\alpha\beta}=(\partial_\gamma\,h_\alpha^A)h_{A\beta}
+(\partial_\gamma\,h_\beta^B)h_{\alpha\,B}+h_\alpha^Ah_\beta^Bh_\gamma^C\partial_Cg_{AB}-h_\alpha^Ah_\beta^Bh_\gamma^n\partial_ng_{AB}.\end{equation}
Because of eq.(19) it becomes
\begin{equation}D_\gamma\,h_{\alpha\beta}\equiv\partial_\gamma\,h_{\alpha\beta}+h_\gamma^n\partial_nh_{\alpha\beta}
=h_\alpha^Ah_\beta^Bh_\gamma^C\partial_Cg_{AB}.\end{equation}
Define that
\begin{equation}P_{\alpha\beta}^\gamma=\frac{1}{2}h^{\gamma\delta}(D_\alpha\,h_{\delta\beta}+D_\beta\,h_{\alpha\delta}
-D_\delta\,h_{\alpha\beta})
\equiv\mathbf{\Gamma}_{\alpha\beta}^\gamma+H_{\alpha\beta}^\gamma,\end{equation}
i.e.\begin{equation}H_{\alpha\beta}^\gamma\equiv-\frac{1}{2}h^{\gamma\delta}(N_\alpha\partial_nh_{\delta\beta}
+N_\beta\partial_nh_{\alpha\delta}-N_\delta\partial_nh_{\alpha\beta}).\end{equation}
Insert eq.(22) into eq.(23) we obtain
\begin{eqnarray*}
\lefteqn{P_{\alpha\beta}^\gamma=}\\
  & &{\frac{1}{2}h^{\gamma\delta}h_\delta^Dh_\alpha^Ah_\beta^B(\partial_Ag_{DB}+\partial_Bg_{AD}-\partial_Dg_{AB})=}\\
  & &{h_C^\gamma\,h_\alpha^Ah_\beta^B\frac{1}{2}g^{CD}(\partial_Ag_{DB}+\partial_Bg_{AD}-\partial_Dg_{AB})=}\\
  & &{h_C^\gamma\,h_\alpha^Ah_\beta^B\Gamma_{AB}^C=}\end{eqnarray*}\begin{eqnarray}
  & &{h_C^\gamma(h_\alpha^Ah_\beta^B\Gamma_{AB}^C+\partial_\alpha\,h_\beta^C)}.\end{eqnarray}
In last step we have used the ambiguity because of eq.(19). It is
clear that
\begin{equation}h_C^\gamma\tilde{\nabla}_\alpha\,h_\beta^C
\equiv\,h_C^\gamma(\partial_\alpha\,h_\beta^C+\Gamma_{AB}^Ch_\alpha^Ah_\beta^B-P_{\alpha\beta}^\delta\,h_\delta^C)=0,\end{equation}
which tells us that $\tilde{\nabla}_\alpha\,h_\beta^C$ is
proportional to the normal vector field $n^C$, hence we can define
a tensor $K_{\alpha\beta}$ by
\begin{equation}\tilde{\nabla}_\alpha\,h_\beta^C=K_{\alpha\beta}n^C\end{equation}
that is the Gauss formula in present case. Operator
$\tilde{\nabla}_\alpha$ we have introduced is an operator which
operates on the n-dimensional index as well as m-dimensional index
simultaneously\cite{Yano}. In fact if we define the following
operations
\begin{equation}\tilde{\nabla}_\beta\,n^C=\partial_\beta\,n^C+\Gamma_{BA}^Ch_\beta^Bn^A,
\qquad\tilde{\nabla}_\beta\,n_C=\partial_\beta\,n_C-\Gamma_{BC}^Ah_\beta^Bn_A;\end{equation}
and
\begin{equation}\tilde{\nabla}_\beta\,u_\gamma=\partial_\beta\,u_\gamma-P_{\beta\gamma}^\alpha\,u_\alpha,\qquad
\tilde{\nabla}_\beta\,u^\gamma=\partial_\beta\,u^\gamma+P_{\beta\alpha}^\gamma\,u^\alpha.\end{equation}
Operator $\tilde{\nabla}_\alpha$ on $h_\beta^C$ certainly agrees
with eq.(26). However we have to note that
\begin{eqnarray*}\lefteqn{\tilde{\nabla}_\beta\,g_{AB}=}\\
  & &{\partial_\beta\,g_{AB}-\Gamma_{DA}^Ch_\beta^Dg_{CB}-\Gamma_{DB}^Ch_\beta^Dg_{CA}=}\end{eqnarray*}\begin{eqnarray}
  & &{-\Gamma_{nA}^Ch_\beta^ng_{CB}-\Gamma_{nB}^Ch_\beta^ng_{CA}\neq{0}}.\end{eqnarray}
From definition eq.(23) and eq.(24) we have
\begin{eqnarray*}\lefteqn{\tilde{\nabla}_\gamma\,h_{\alpha\beta}=}\\
  & &{\partial_\gamma\,h_{\alpha\beta}-P_{\gamma\beta}^\delta\,h_{\alpha\delta}-P_{\alpha\gamma}^\delta\,h_{\delta\beta}=}\\
  & &{\mathbf{\nabla}_\gamma\,h_{\alpha\beta}-H_{\gamma\beta}^\delta\,h_{\alpha\delta}-H_{\alpha\gamma}^\delta\,h_{\delta\beta}=}\end{eqnarray*}
\begin{eqnarray}
  & &{-h_\gamma^n\partial_nh_{\alpha\beta}}\end{eqnarray}\\
or
\begin{eqnarray*}\lefteqn{\tilde{\nabla}_\gamma\,h_{\alpha\beta}=}\\
  & &{h_\alpha^Ah_\beta^Bh_\gamma^C\partial_Cg_{AB}-h_C^\delta\,h\gamma^Ah_\beta^B\Gamma_{AB}^Ch_{\alpha\delta}-}\\
  & &{h_\alpha^Ah_\gamma^C\Gamma_{AC}^Dh_\beta^B(g_{DB}-\epsilon\,n_D\,n_B)-h_\gamma^n\partial_nh_{\alpha\beta}=}\\
  & &{h_\alpha^Ah_\beta^Bh_\gamma^C(\partial_Cg_{AB}-\Gamma_{CA}^Dg_{DB}-\Gamma_{CB}^Dg_{AD})-h_\gamma^n\partial_nh_{\alpha\beta}=}\\
  & &{h_\alpha^Ah_\beta^Bh_\gamma^C\nabla_Cg_{AB}-h_\gamma^n\partial_nh_{\alpha\beta}=}\end{eqnarray*}\begin{eqnarray}
  & &{-h_\gamma^n\partial_nh_{\alpha\beta}}.\end{eqnarray}
Eq.(25)leads to
\begin{eqnarray*}\lefteqn{\tilde{\nabla}_\alpha\,h_\beta^C=}\\
  & &{\partial_\alpha\,h_\beta^C+\Gamma_{AB}^Ch_\alpha^Ah_\beta^B-P_{\alpha\beta}^\gamma\,h_\gamma^C=}\\
  & &{(\delta_D^C-h_D^\gamma\,h_\gamma^C)(\partial_\alpha\,h_\beta^D+\Gamma_{AB}^Dh_\alpha^Ah_\beta^B)=}\end{eqnarray*}\begin{eqnarray}
  & &{\epsilon\,n_D(\partial_\alpha\,h_\beta^D+\Gamma_{AB}^Dh_\alpha^Ah_\beta^B)n^C}
\end{eqnarray} therefore
\begin{eqnarray*}\lefteqn{K_{\alpha\beta}=}\\
  & &{\epsilon\,n_D(\partial_\alpha\,h_\beta^D+\Gamma_{AB}^Dh_\alpha^Ah_\beta^B)=}\\
  & &{-\frac{\epsilon}{N}\partial_\alpha\,N_\beta+\frac{\epsilon}{N}(\Gamma_{AB}^n+N_\gamma\Gamma_{AB}^\gamma)h_\alpha^Ah_\beta^B=}\\
  & &{-\frac{\epsilon}{N}\partial_\alpha\,N_\beta+\frac{\epsilon}{N}(\Gamma_{\alpha\beta}^n+N_\gamma\Gamma_{\alpha\beta}^\gamma)}\\
  & &{-\frac{\epsilon\,N_\alpha}{N}(\Gamma_{\beta\,n}^n+N_\gamma\Gamma_{\beta\,n}^\gamma)
     -\frac{\epsilon\,N_\beta}{N}(\Gamma_{\alpha\,n}^n}\\
  & &{+N_\gamma\Gamma_{\alpha\,n}^\gamma)+\frac{\epsilon}{N}N_\alpha\,N_\beta(\Gamma_{nn}^n+N_\gamma\Gamma_{nn}^\gamma)=}\end{eqnarray*}\begin{eqnarray}
  & &{-\frac{\epsilon}{2N}\mathcal{F}_{\alpha\beta}-\frac{1}{2N}\mathcal{H}_{\alpha\beta}}.\end{eqnarray}
in which
\begin{equation}\mathcal{F}_{\alpha\beta}\equiv\partial_\alpha\,N_\beta-\partial_\beta\,N_\alpha,\qquad
\mathcal{H}_{\alpha\beta}\equiv\,N^2\partial_nh_{\alpha\beta}+\epsilon\partial_n(N_\alpha\,N_\beta).\end{equation}
Similar to subsection {Lapse review} we can also define
\begin{equation}K_{AB}=h_A^\alpha\,h_B^\beta\,K_{\alpha\beta},\qquad
K=K_A^A=K_\alpha^\alpha.\end{equation} Here we want to point out
that the tensor $K_{AB}$ has not only a symmetric part
($\mathcal{H}$) but also an antisymmetric part ($\mathcal{F}$).
Since\begin{eqnarray}0=\tilde{\nabla}_\beta(n^An_A)&=&(\tilde{\nabla}_\beta\,n^A)n_A+n^A\tilde{\nabla}_\beta\,n_A\\
&=&n_Bn_A\tilde{\nabla}_\beta\,g^{AB}+2n^A\tilde{\nabla}_\beta\,n_A\\&=&2\Gamma_{nB}^Ah_\beta^nn^Bn_A+2n^A\tilde{\nabla}_\beta\,n_A\end{eqnarray}
so that
\begin{equation}n^A\tilde{\nabla}_\beta\,n_A=-\Gamma_{nB}^Ah_\beta^nn^Bn_A,\end{equation}
we then obtain the Weingarten formula as following
\begin{eqnarray*}\lefteqn{\tilde{\nabla}_\beta\,n_A=}\\
  & &{h_A^B\tilde{\nabla}_\beta\,n_B-\epsilon\,n_A\Gamma_{nC}^Bh_\beta^nn^Cn_B=}\\
  & &{h_A^\alpha\,h_\alpha^B(\partial_\beta\,n_B-\Gamma_{BC}^Dh_\beta^Cn_D)-\epsilon\,n_A\Gamma_{nC}^Bh_\beta^nn^Cn_B=}\\
  & &{-h_A^\alpha(\partial_\beta\,h_\alpha^D+\Gamma_{BC}^Dh_\alpha^Bh_\beta^C)n_D-\epsilon\,n_A\Gamma_{nC}^Bh_\beta\,^nn^Cn_B=}\end{eqnarray*}\begin{eqnarray}
  & &{-\epsilon\,h_A^\alpha\,K_{\beta\alpha}-\epsilon\Gamma_{nC}^Bh_\beta^nn^Cn_Bn_A},\end{eqnarray}
or
\begin{eqnarray*}\lefteqn{\tilde{\nabla}_\beta\,n^A=}\\
  & &{-\epsilon\,h_\alpha^AK_\beta{}^\alpha+(\Gamma_{nB}^An^B+\Gamma_{nB}^Ch^{AB}n_C)h_\beta^n=}\end{eqnarray*}\begin{eqnarray}
  & &{-\epsilon\,h_\alpha^AK_\beta{}^\alpha+(\frac{1}{N}h^{A\alpha}\partial_nN_\alpha+\mathbf{b}n^A\partial_n\phi)h_\beta^n}.\end{eqnarray}
\section{Variation of Gauss Equation}
To get the Gauss equation we have to calculate
$\tilde{\nabla}_\gamma\tilde{\nabla}_\alpha\,h_\beta^C$, that is
\begin{eqnarray*}\lefteqn{\tilde{\nabla}_\gamma\tilde{\nabla}_\alpha\,h_\beta^C=}\\
  & &{\partial_\gamma\partial_\alpha\,h_\beta^C+(\partial_\alpha\Gamma_{AB}^C)h_\alpha^Ah_\beta^Bh_\gamma^D-}\\
  & &{(\partial_n\Gamma_{AB}^C)h_\alpha^Ah_\beta^Bh_\gamma^n+\Gamma_{AB}^C(\partial_\gamma\,h_\alpha^A)h_\beta^B}\\
  & &{+\Gamma_{AB}^Ch_\alpha^A\partial_\gamma\,h_\beta^B-(\partial_\gamma\,P_{\alpha\beta}^\delta)h_\delta^C-}\\
  & &{P_{\alpha\beta}^\delta\partial_\gamma\,h_\delta^C+\Gamma_{DF}^Ch_\gamma^D(\partial_\alpha\,h_\beta^F+}\\
  & &{\Gamma_{AB}^Fh_\alpha^Ah_\beta^B-P_{\alpha\beta}^\delta\,h_\delta^F)-(\partial_\alpha\,h_\delta^C+}\\
  & &{\Gamma_{AB}^Ch_\alpha^Ah_\delta^B-P_{\alpha\delta}^\eta\,h_\eta^C)P_{\gamma\beta}^\delta-(\partial_\delta\,h_\beta^C+}\end{eqnarray*}\begin{eqnarray}
  & &\Gamma_{AB}^Ch_\delta^Ah_\beta^B-P_{\delta\beta}^\eta\,h_\eta^C)P_{\gamma\alpha}^\delta.\end{eqnarray}
From the definition of Riemann tensor we find
\begin{eqnarray*}\lefteqn{(\tilde{\nabla}_\gamma\tilde{\nabla}_\alpha-\tilde{\nabla}_\alpha\tilde{\nabla}_\gamma)h_\beta^C=}\\
  & &{h_\alpha^Ah_\beta^Bh_\gamma^D(\partial_D\Gamma_{AB}^C-\partial_A\Gamma_{DB}^C+\Gamma_{DF}^C\Gamma_{AB}^F-
     \Gamma_{AF}^C\Gamma_{DB}^F)-h_\delta^C(\partial_\gamma\,P_{\alpha\beta}^\delta-\partial_\alpha\,P_{\gamma\beta}^\delta-}\\
  & &{P_{\alpha\eta}^\delta\,P_{\gamma\beta}^\eta+P_{\gamma\eta}^\delta\,P_{\alpha\beta}^\eta)-
     (\partial_n\Gamma_{AB}^C)h_\beta^B(h_\alpha^Ah_\gamma^n-h_\gamma^Ah_\alpha^n)+\Gamma_{AB}^Ch_\beta^B(\partial_\gamma\,h_\alpha^A-}\\
  & &{\partial_\alpha\,h_\gamma^A)=h_\alpha^Ah_\beta^Bh_\gamma^D{R_{ADB}}^C-h_\delta^C{S_{\alpha\gamma\beta}}^\delta-
     (\partial_n\Gamma_{\alpha\,B}^Ch_\gamma^n-\partial_n\Gamma_{\gamma\,B}^Ch_\alpha^n)h_\beta^B+}\end{eqnarray*}\begin{eqnarray}
  & &{\Gamma_{nB}^Ch_\beta^B(\partial_\gamma\,h_\alpha^n-\partial_\alpha\,h_\gamma^n)}.\end{eqnarray}
In which we have used a symbol\\
\begin{eqnarray*}\lefteqn{S_{\alpha\gamma\beta}{}^\delta\equiv\partial_\gamma\,P_{\alpha\beta}^\delta-\partial_\alpha\,P_{\gamma\beta}^\delta-
     P_{\alpha\eta}^\delta\,P_{\gamma\beta}^\eta+P_{\gamma\eta}^\delta\,P_{\alpha\beta}^\eta=}\\
  & &{\mathbf{{R}_{\alpha\gamma\beta}}^\delta+\partial_\gamma\,H_{\alpha\beta}^\delta-\partial_\alpha\,H_{\gamma\beta}^\delta+
     H_{\gamma\eta}^\delta\,H_{\alpha\beta}^\eta-H_{\alpha\eta}^\delta\,H_{\gamma\beta}^\eta+}\end{eqnarray*}\begin{eqnarray}
  & &{\Gamma_{\gamma\eta}^\delta\,H_{\alpha\beta}^\eta+\Gamma_{\alpha\beta}^\eta\,H_{\gamma\eta}^\delta-
     \Gamma_{\alpha\eta}^\delta\,H_{\gamma\beta}^\eta-\Gamma_{\gamma\beta}^\eta\,H_{\alpha\eta}^\delta}.\end{eqnarray}\\
Tensor ${S_{\alpha\gamma\beta}}^\delta$ has the following
symmetric properties like Riemann tensor
\begin{equation}{S_{\alpha\gamma\beta}}^\delta=-{S_{\gamma\alpha
\beta}}^\delta \end{equation} and
\begin{equation}{S_{\alpha\gamma\beta}}^\delta+{S_{\gamma\beta\alpha}}^\delta+{S_{\beta\alpha\gamma}}^\delta=0
\end{equation}
 but have no corresponding symmetries after contracting with a metric tensor.
 It is easy to find also
 \begin{equation}\tilde{\nabla}_\alpha\,h_B^\gamma=K_\alpha{}^\gamma\,n_B-h_{B\beta}h_\alpha^n\partial_nh^{\gamma\beta}-
 h^{C\gamma}\Gamma_{nC}^Ah_\alpha^ng_{AB}-h_A^\gamma\Gamma_{nB}^Ah_\alpha^n.\end{equation}
 As a result we obtain
 \begin{eqnarray*}\lefteqn{h_C^\zeta\,h_\alpha^Ah_\beta^Bh_\gamma^D{R_{ADB}}^C-{S_{\alpha\gamma\beta}}^\zeta+
      h_C^\zeta\,h_\beta^B\Gamma_{nB}^C(\partial_\gamma\,h_\alpha^n-\partial_\alpha\,h_\gamma^n)
      -h_C^\zeta\,h_\beta^B(h_\gamma^n\partial_n\Gamma_{\alpha\,B}^C-h_\alpha^n\partial_n\Gamma_{\gamma\,B}^C)=}\\
   & &{h_C^\zeta(\tilde{\nabla}_\gamma\tilde{\nabla}_\alpha-\tilde{\nabla}_\alpha\tilde{\nabla}_\gamma)h_\beta^C=}\\
   & &{\tilde{\nabla}_\alpha\,h_C^\zeta\tilde{\nabla}_\gamma\,h_\beta^C-
      \tilde{\nabla}_\gamma\,h_C^\zeta\tilde{\nabla}_\alpha\,h_\beta^C=}\\
   & &{K_{\gamma\beta}n^C(K_\alpha{}^\zeta\,n_C-h_{C\delta}h_\alpha^n\partial_nh^{\zeta\delta}-
      h^{B\zeta}\Gamma_{nB}^Ah_\alpha^ng_{AC}-h_A^\zeta\Gamma_{nC}^Ah_\alpha^n)-K_{\alpha\beta}n^C(K_\gamma{}^\zeta\,n_C-}\\
   & &{h_{C\delta}h_\gamma^n\partial_nh^{\zeta\delta}-h^{B\zeta}\Gamma_{nB}^Ah_\gamma^ng_{AC}-h_A^\zeta\Gamma_{nC}^Ah_\gamma^n)=}\end{eqnarray*}\begin{eqnarray}
   & &{\epsilon\,K_{\gamma\beta}K_\alpha{}^\zeta-\epsilon\,K_{\alpha\beta}K_\gamma{}^\zeta-
      (h_\alpha^nK_{\gamma\beta}-h_\gamma^nK_{\alpha\beta})(h_A^\zeta\,n^B+h^{B\zeta}n_A)\Gamma_{nB}^A}.\end{eqnarray}
Contracting $\gamma$ with $\zeta$, and multiplying
$h^{\alpha\beta}$, then eq.(49) becomes
\begin{eqnarray*}\lefteqn{h^{AB}h^{CD}R_{ADBC}=R-2\epsilon\,R_{ab}n^an^b=R-2\epsilon\,N^2R_{nn}=}\\
  & &{\mathbf{R}+h^{\alpha\beta}\partial_\gamma\,H_{\alpha\beta}^\gamma-h^{\alpha\beta}\partial_\alpha\,H_{\gamma\beta}^\gamma+
     \mathbf{\Gamma}_{\gamma\eta}^\gamma\,h^{\alpha\beta}H_{\alpha\beta}^\eta-}\\
  & &{2h^{\alpha\beta}\mathbf{\Gamma}_{\alpha\eta}^\gamma\,H_{\gamma\beta}^\eta+h^{\alpha\beta}\mathbf{\Gamma}_{\alpha\beta}^\eta\,H_{\gamma\eta}^\gamma+
     H_{\gamma\eta}^\gamma\,h^{\alpha\beta}H_{\alpha\beta}^\eta-}\\
  & &{h^{\alpha\beta}H_{\alpha\eta}^\gamma\,H_{\gamma\beta}^\eta-h^{B\alpha}\Gamma_{nB}^\gamma\mathcal{F}_{\alpha\gamma}+
     h^{B\alpha}(N_\alpha\partial_n\Gamma_{\gamma\,B}^\gamma-N_\gamma\partial_n\Gamma_{\alpha\,B}^\gamma)-}\\
  & &{\frac{\epsilon}{4N^2}\mathcal{F}_{\alpha\gamma}\mathcal{F}^{\alpha\gamma}+
     \frac{\epsilon}{4N^2}(\mathcal{H}_{\alpha\gamma}\mathcal{H}^{\alpha\gamma}-\mathcal{H}^2)-}\\
  & &{\frac{1}{2N^2}[N_\alpha(\epsilon\mathcal{F}^{\beta\alpha}+\mathcal{H}^{\beta\alpha})-N^\beta\mathcal{H}]\partial_nN_\beta=}\\
  & &{\mathbf{R}-\frac{3\epsilon}{4N^2}\mathcal{F}_{\alpha\gamma}\mathcal{F}^{\alpha\gamma}+
     \frac{3\epsilon}{N^2}\mathcal{F}^{\alpha\gamma}N_\alpha\partial_nN_\gamma+N_\alpha\partial^\gamma\,h^{\alpha\beta}\partial_nh_{\beta\gamma}+}\\
  & &{h^{\alpha\beta}(\partial^\gamma\,N_\gamma\partial_nh_{\alpha\beta}-
     \partial^\gamma\,N_\alpha\partial_nh_{\beta\gamma})+N^\alpha\,h^{\gamma\delta}\partial_\alpha\partial_nh_{\gamma\delta}+}\\
  & &{\frac{\epsilon}{4N^2}(\mathcal{H}_{\alpha\gamma}\mathcal{H}^{\alpha\gamma}-\mathcal{H}^2)+
     \frac{1}{N^2}(N^\alpha\,N_\gamma\partial_n\mathcal{H}_\alpha^\gamma-N_\alpha\,N^\alpha\partial_n\mathcal{H})-}\\
  & &{\frac{1}{2N^2}(N_\alpha\,N^\alpha\mathcal{H}^{\gamma\delta}-N^\gamma\,N^\delta\mathcal{H})\partial_nh_{\gamma\delta}+
     \frac{2\mathbf{b}}{N^2}(N^\alpha\,N^\gamma\mathcal{H}_{\alpha\gamma}-N_\gamma\,N^\gamma\mathcal{H})\partial_n\phi+}\end{eqnarray*}\begin{eqnarray}
  & &{\frac{1}{4}h^{\alpha\beta}(2N^\gamma\,N^\delta-h^{\gamma\delta}N_\lambda\,N^\lambda)(\partial_nh_{\alpha\beta}\partial_nh_{\gamma\delta}-
     \partial_nh_{\alpha\delta}\partial_nh_{\beta\gamma})}\end{eqnarray}
in which we have used n-1 dimensional harmonic condition
$h^{\alpha\beta}\mathbf{\Gamma}_{\alpha\beta}^\gamma=0$. On the
other hand, we can calculate the second term on the left hand side
of the equation
\begin{eqnarray*}\lefteqn{2\epsilon\,R_{ab}n^an^b=}\\
  & &{\frac{\epsilon}{2N^2}\mathcal{F}_{\alpha\beta}\mathcal{F}^{\alpha\beta}-
     \frac{2\epsilon}{N^2}\mathcal{F}^{\alpha\beta}N_\alpha\partial_nN_\beta-}\\
  & &{\frac{\epsilon}{2N^2}(\mathcal{H}_{\alpha\beta}\mathcal{H}^{\alpha\beta}-\mathcal{H}^2)+
     \frac{2}{N^2}\mathcal{H}^{\alpha\beta}\,N_\alpha\partial_nN_\beta-\frac{2}{N^2}\mathcal{H}N^\alpha\partial_nN_\alpha+}\end{eqnarray*}\begin{eqnarray}
  & &{2\epsilon[\nabla_C(n^A\nabla_An^C)-\nabla_A(n^A\nabla_Cn^C)]}.\end{eqnarray}
Therefore\\
\begin{eqnarray*}
R&=&{\mathbf{R}-\frac{\epsilon}{4N^2}\mathcal{F}_{\alpha\beta}\mathcal{F}^{\alpha\beta}+
    \frac{\epsilon}{N^2}\mathcal{F}^{\alpha\beta}N_\alpha\partial_nN_\beta}\\
 & &{+N_\alpha\partial^\gamma\,h^{\alpha\beta}\partial_nh_{\beta\gamma}+
    h^{\alpha\beta}(\partial^\gamma\,N_\gamma\partial_nh_{\alpha\beta}}\\
 & &{-\partial^\gamma\,N_\alpha\partial_nh_{\beta\gamma})+
    N^\alpha\,h^{\gamma\delta}\partial_\alpha\partial_nh_{\gamma\delta}}\\
 & &{+\frac{3\epsilon}{4N^2}(\mathcal{H}_{\alpha\beta}\mathcal{H}^{\alpha\beta}-\mathcal{H}^2)
    -\frac{1}{2N^2}(N_\gamma\,N^\gamma+2\epsilon\,N^2)\mathcal{H}^{\alpha\beta}\partial_nh_{\alpha\beta}+}\\
 & &{\frac{\epsilon}{2N^2}(N^\alpha\,N^\gamma+2N^2h^{\alpha\gamma})\mathcal{H}\partial_nh_{\alpha\beta}+
    \frac{1}{N^2}(N^\alpha\,N_\gamma\partial_n\mathcal{H}_\alpha^\gamma}\\
 & &{-N_\alpha\,N^\alpha\partial_n\mathcal{H})+\frac{2\mathbf{b}}{N^2}(N^\alpha\,N^\gamma\mathcal{H}_{\alpha\gamma}
    -N_\alpha\,N^\alpha\mathcal{H})\partial_n\phi+}\\
 & &{\frac{1}{4}h^{\alpha\beta}(2N^\gamma\,N^\delta-h^{\gamma\delta}N_\lambda\,N^\lambda)(\partial_nh_{\alpha\beta}\partial_nh_{\gamma\delta}-}\end{eqnarray*}\begin{eqnarray}
 & &{\partial_nh_{\alpha\delta}\partial_nh_{\beta\gamma})+2\epsilon[\nabla_C(n^A\nabla_An^C)-\nabla_A(n^A\nabla_Cn^C)]}.\end{eqnarray}
The last divergence term in above equation, which may be neglected
in Hamiltonian formulation \cite{Weinberg}\cite{Wald}, is
important in the dimensional reduction. So we now evaluate it as
well\\
\begin{eqnarray*}\lefteqn{\nabla_C(n^A\nabla_An^C)-\nabla_A(n^A\nabla_Cn^C)=}\\
  & &{\epsilon\nabla_C[N(\partial_nn^C+\epsilon\,N\Gamma_{nn}^C)]-\epsilon\,N\nabla_Cn^C\Gamma_{\alpha\,n}^\alpha-
     \epsilon\partial_n(N\nabla_Cn^C)-\epsilon\,N\nabla_Cn^C\Gamma_{nn}^n=}\\
  & &{\partial_\alpha(N^2\Gamma_{nn}^\alpha)+\partial_n(N\partial_nN+N^2\Gamma_{nn}^n)+\partial_n[-\mathcal{H}/2+
     N^2N_\alpha\Gamma_{nn}^\alpha+\epsilon\,N^\beta(\Gamma_{n\beta}^n}\\
  & &{+N_\alpha\Gamma_{n\beta}^\alpha)-\epsilon\,N_\gamma\,N^\gamma(\Gamma_{nn}^n+N_\alpha\Gamma_{nn}^\alpha)]
     +(N\partial_nN+N^2\Gamma_{nn}^n)(\Gamma_{\alpha\,n}^\alpha+}\\
  & &{\Gamma_{nn}^n)+N^2(\Gamma_{\alpha\beta}^\alpha\Gamma_{nn}^\beta+\Gamma_{n\beta}^n\Gamma_{nn}^\beta)+
     [-\mathcal{H}/2+N^2N_\alpha\Gamma_{nn}^\alpha+\epsilon\,N_\beta(\Gamma_{n\beta}^n+}\end{eqnarray*}\begin{eqnarray}
  & &{N_\alpha\Gamma_{n\beta}^\alpha)-\epsilon\,N_\gamma\,N^\gamma(\Gamma_{nn}^n+
     N_\alpha\Gamma_{nn}^\alpha)](\Gamma_{\alpha\,n}^\alpha+\Gamma_{nn}^n)}. \end{eqnarray} \\
From Appendix we easily find
\begin{equation}\Gamma_{\alpha\beta}^\alpha+\Gamma_{n\beta}^n=\mathbf{\Gamma}_{\alpha\beta}^\alpha+\mathbf{b}\partial_\beta\phi
\qquad\Gamma_{\alpha\,n}^\alpha+\Gamma_{nn}^n=\mathbf{b}\partial_\beta\phi+\frac{1}{2}h^{\alpha\beta}\partial_nh_{\alpha\beta}\end{equation}
\begin{equation}\Gamma_{n\beta}^n+N_\alpha\Gamma_{n\beta}^\alpha=\mathbf{b}\partial_\beta\phi\qquad
\Gamma_{nn}^n+N_\alpha\Gamma_{nn}^\alpha=-\frac{1}{N}\partial_nN=\mathbf{b}\partial_n\phi.\end{equation}
Eventually we obtain\\
\begin{eqnarray*}\lefteqn{\nabla_C(n^A\nabla_An^C)-\nabla_A(n^A\nabla_Cn^C)=}\\
  & &{-\epsilon\mathbf{b}(\partial^\alpha\partial_\alpha\phi+\mathbf{b}\partial^\alpha\phi\partial_\alpha\phi)+
     \epsilon\partial^\alpha\partial_nN_\alpha+\epsilon\mathbf{b}(2\partial^\alpha\phi\partial_nN_\alpha+\partial^\alpha\,N_\alpha\partial_n\phi}\\
  & &{+2N^\alpha\partial_\alpha\partial_n\phi+2\mathbf{b}N^\alpha\partial_\alpha\phi\partial_n\phi-N^\alpha\partial^\beta\phi\partial_nh_{\alpha\beta}
     +\frac{1}{2}N^\gamma\partial_\gamma\phi\,h^{\alpha\beta}\partial_nh_{\alpha\beta})}\\
  & &{-\partial_n(\mathcal{H}/2+\epsilon\mathbf{b}N_\gamma\,N^\gamma\partial_n\phi)-(\mathcal{H}/2+}\end{eqnarray*}\begin{eqnarray}
  & &{\epsilon\mathbf{b}N_\gamma\,N^\gamma\partial_n\phi)(\mathbf{b}\partial_n\phi
     +\frac{1}{2}h^{\alpha\beta}\partial_nh_{\alpha\beta})}\end{eqnarray}\\
in which n-1 dimensional harmonic condition is also used.
\section{Variation of Codazzi Equation}
First of all, let us calculate the double covariant derivative of
normal vector $n^C$
\begin{eqnarray*}\lefteqn{\tilde{\nabla}_\gamma\tilde{\nabla}_\beta\,n^C=}\\
  & &{\partial_\gamma\partial_\beta\,n^C+(\partial_D\Gamma_{BA}^C)h_\beta^Bh_\gamma^Dn^A-
     (\partial_n\Gamma_{BA}^C)h_\beta^Bh_\gamma^nn^A+\Gamma_{BA}^C(\partial_\gamma\,h_\beta^B)n^A}\\
  & &{+\Gamma_{BA}^Ch_\beta^B\partial_\gamma\,n^A+\Gamma_{DA}^Ch_\gamma^D\partial_\beta\,n^A
     +\Gamma_{DA}^C\Gamma_{BF}^Ah_\gamma^Dh_\beta^Bn^F-P_{\gamma\beta}^\alpha\partial_\alpha\,n^C-}\end{eqnarray*}\begin{eqnarray}
  & &{P_{\gamma\beta}^\alpha\Gamma_{AF}^Ch_\alpha^An^F}.\end{eqnarray} Next,
we have to antisymmetrize it as follows
\begin{eqnarray*}\lefteqn{(\tilde{\nabla}_\gamma\tilde{\nabla}_\beta-\tilde{\nabla}_\beta\tilde{\nabla}_\gamma)n^C=}\\
  & &{(\partial_D\Gamma_{BA}^C-\partial_B\Gamma_{DA}^C)h_\beta^Bh\gamma^Dn^A+(\Gamma_{DF}^C\Gamma_{BA}^F-
     \Gamma_{BF}^C\Gamma_{DA}^F)h_\gamma^Dh_\beta^Bn^A}\\
  & &{+\Gamma_{BA}^C(\partial_\gamma\,h_\beta^B-\partial_\beta\,h_\gamma^B)n^A-(\partial_n\Gamma_{BA}^C)(h_\beta^Bh_\gamma^n-
     h_\gamma^Bh_\beta^n)n^A=}\\
  & &{{R_{BDA}}^Ch_\beta^Bh\gamma^Dn^A+\epsilon\,N(\partial_\gamma\,h_\beta^n-\partial_\beta\,h_\gamma^n)\Gamma_{nn}^C
     -(\partial_n\Gamma_{BA}^C)(h_\beta^Bh_\gamma^n}\end{eqnarray*}{\begin{eqnarray}
  & &{-h_\gamma^Bh_\beta^n)n^A}.\end{eqnarray}
By using of the Weingarten formula (42) the left hand side of
eq.(58) becomes\\
\begin{eqnarray*}\lefteqn{(\tilde{\nabla}_\gamma\tilde{\nabla}_\beta-\tilde{\nabla}_\beta\tilde{\nabla}_\gamma)n^C=}\\
  & &{-\epsilon\,h_\alpha^C(\tilde{\nabla}_\gamma\,K_\beta{}^\alpha-\tilde{\nabla}_\beta\,K_\gamma{}^\alpha)-
     \epsilon(K_\beta{}^\alpha\,K_{\gamma\alpha}-K_\gamma{}^\alpha\,K_{\beta\alpha})n^C+(\frac{1}{N}h^{C\alpha}\partial_nN_\alpha}\\
  & &{+\mathbf{b}n^C\partial_n\phi)(\partial_\gamma\,h_\beta^n-\partial_\beta\,h_\gamma^n)+\partial_\gamma(\frac{1}{N}h^{C\alpha}\partial_nN_\alpha
     +\mathbf{b}n^C\partial_n\phi)h_\beta^n-\partial_\beta(\frac{1}{N}h^{C\alpha}\partial_nN_\alpha}\end{eqnarray*}\begin{eqnarray}
  & &{+\mathbf{b}n^C\partial_n\phi)h_\gamma^n+(\Gamma_{AD}^Ch_\gamma^Ah_\beta^n-\Gamma_{AD}^Ch_\beta^Ah_\gamma^n)(\frac{1}{N}h^{D\alpha}\partial_nN_\alpha
     +\mathbf{b}n^D\partial_n\phi)}\end{eqnarray}\\
multiplying $h_C^\alpha$ on both sides of eq.(58), we would see
\begin{eqnarray*}\lefteqn{-\epsilon(\tilde{\nabla}_\gamma\,{K_\beta}^\alpha-\tilde{\nabla}_\beta\,{K_\gamma}^\alpha)=}\\
  & &{{R_{BDA}}^Ch_C^\alpha\,h_\beta^Bh_\gamma^Dn^A+\epsilon\,Nh_C^\alpha(\partial_\gamma\,h_\beta^n-
     \partial_\beta\,h_\gamma^n)\Gamma_{nn}^C+\epsilon\,Nh_C^\alpha(h_\beta^n\partial_n\Gamma_{\gamma\,n}^C}\\
  & &{-h_\gamma^n\partial_n\Gamma_{\beta\,n}^C)-\frac{1}{N}h^{\alpha\delta}\partial_nN_\delta(\partial_\gamma\,h_\beta^n-
     \partial_\beta\,h_\gamma^n)-\partial_\gamma(\frac{1}{N}h^{\alpha\delta}\partial_nN_\delta)h_\beta^n}\\
  & &{+\partial_\beta(\frac{1}{N}h^{\alpha\delta}\partial_nN_\delta)h_\gamma^n-h_C^\alpha\Gamma_{AD}^C(h_\gamma^Ah_\beta^n-}\end{eqnarray*}\begin{eqnarray}
  & &{h_\beta^Ah_\gamma^n)(\frac{1}{N}h^{D\delta}\partial_nN_\delta+\mathbf{b}n^D\partial_n\phi)}.\end{eqnarray}
Contracting index $\gamma$ with $\alpha$ we achieve the goal.\\
\begin{eqnarray*}\lefteqn{-\epsilon(\tilde{\nabla}_\alpha\,K_\beta^\alpha-\tilde{\nabla}_\beta\,K)=}\\
  & &{R_{BA}h_\beta^Bn^A+\epsilon\,N(\partial_\alpha\,h_\beta^n-\partial_\beta\,h_\gamma^n)\Gamma_{nn}^\alpha+
     \epsilon\,N(h_\beta^n\partial_n\Gamma_{\alpha\,n}^\alpha-h_\alpha^n\partial_n\Gamma_{\beta\,n}^\alpha)}\\
  & &{-\frac{1}{N}h^{\gamma\alpha}\partial_nN_\alpha(\partial_\gamma\,h_\beta^n-\partial_\beta\,h_\gamma^n)-
     \partial_\alpha(\frac{1}{N}h^{\alpha\gamma}\partial_nN_\gamma)h_\beta^n+}\\
  & &{\partial_\beta(\frac{1}{N}h^{\alpha\gamma}\partial_nN_\gamma)h_\alpha^n-(\Gamma_{\alpha\gamma}^\alpha\,h_\beta^n-
     \Gamma_{\beta\gamma}^\alpha\,h_\alpha^n)(\frac{1}{N}h^{\gamma\delta}\partial_nN_\delta)}\\
  & &{+(\Gamma_{\alpha\,n}^\alpha\,h_\beta^n-\Gamma_{\beta\,n}^\alpha\,h_\alpha^n)(\frac{1}{N}N^\gamma\partial_nN_\gamma-
     \epsilon\mathbf{b}N\partial_n\phi)=}\\
  & &{R_{BA}h_\beta^Bn^A+\frac{\mathbf{b}}{N}\mathcal{F}_{\alpha\beta}\partial^\alpha\phi-
     \frac{3\mathbf{b}}{2N}N^\alpha\mathcal{F}_{\alpha\beta}\partial_n\phi}\\
  & &{+\frac{1}{2N}N_\alpha\mathcal{F}_{\beta\gamma}\partial_nh^{\alpha\gamma}-
     \frac{1}{2N}N^\gamma(\partial_\beta\partial_nN_\gamma+\partial_\gamma\partial_nN_\beta)+}\\
  & &{\frac{1}{N}h^{\alpha\gamma}N_\beta\partial_\alpha\partial_nN_\gamma
     -\frac{\mathbf{b}}{N}(N^\alpha\partial_\alpha\phi\partial_nN_\beta}\\
  & &{+\partial_\beta\phi\,N^\alpha\partial_nN_\alpha-2N_\beta\partial^\alpha\phi\partial_nN_\alpha)
     -\frac{1}{N}N_\alpha(\partial_\beta\,h^{\alpha\delta}+h^{\gamma\delta}\mathbf{\Gamma}_{\beta\gamma}^\alpha)\partial_nN_\delta}\\
  & &{+\frac{\epsilon}{2N}(N^\alpha\partial_n\mathcal{H}_{\alpha\beta}-N_\beta\,h^{\alpha\gamma}\partial_n\mathcal{H}_{\alpha\gamma})}\\
  & &{+\frac{\epsilon}{4N^3}(N^\gamma\mathcal{H}_{\beta\gamma}-N_\beta\mathcal{H})(\mathcal{H}-N^2h^{\alpha\delta}\partial_nh_{\alpha\delta})+
     \frac{\epsilon\mathbf{b}}{2N}(N^\alpha\mathcal{H}_{\alpha\beta}-N_\beta\mathcal{H})\partial_n\phi}\\
  & &{+\frac{\mathbf{b}}{N}N^\alpha(N_\alpha\partial_nN_\beta-N_\beta\partial_nN_\alpha)\partial_n\phi}\\
  & &{+\frac{\epsilon}{2N}(N_\alpha\mathcal{H}_{\beta\gamma}-N_\beta\mathcal{H}_{\alpha\gamma})\partial_nh^{\alpha\gamma}}\end{eqnarray*}\begin{eqnarray}
  & &{-\frac{1}{2N^3}N^\alpha(N_\alpha\mathcal{H}_{\beta\gamma}-
     N_\beta\mathcal{H}_{\alpha\gamma})h^{\gamma\delta}\partial_nN_\delta}.\end{eqnarray}.
\section{Concerning Cauchy Problem and Hamiltonian Formulation}
The Cauchy problem in general relativity we are interested in
starts from certain data on an initial n-1 dimensional space-like
surface to look for their subsequent evolution with the aid of
Einstein field equation (for simplicity, in vacuum ). Obviously we
now have $\epsilon=-1$. Einstein equation includes evolution
equation and constraint equation. To solve the field equations we
have to find the constraint equations first. For metric (15,17) we
easily write down the following n constraint equations
\begin{equation}(N_\alpha\,N^\alpha+\epsilon\,N^2)R_{nn}-N^\beta\,R_{\beta\,n}-\frac{1}{2}R=0,\end{equation}\\
\begin{equation}(N_\alpha\,N^\alpha+\epsilon\,N^2)R_{\beta\,n}-N^\alpha\,R_{\alpha\beta}=0\end{equation}
because they involve no second time derivatives. Eq.(62) can be
simply related to Gauss-Codazzi equations. It can be reduced to
\begin{equation}\frac{1}{2}(R-2\epsilon\,R_{ab}n^an^b)+\frac{\epsilon}{N}N^\beta\,R_{ab}h_\beta^an^b=0.\end{equation}
However, the relation between other n-1 constraint equations and
Gauss-Codazzi equations is not so obvious. Besides, in derivation
of Gauss-Codazzi equations we have used only the n-1 dimensional
harmonic conditions. To fix the "gauge" it seems necessary to find
n "gauge"(coordinate) conditions.\\As for the Hamiltonian
formulation, presently due to that not only the time derivative
$\dot{h}_{\alpha\beta}(\equiv\partial_nh_{\alpha\beta})$, but also
$\dot{N}$ or $\dot {\phi}$ and $ \dot{N}_\alpha$ appear in
Lagrangian
\begin{equation}\mathcal{L}=\sqrt{-g}R=\frac{\sqrt{h}}{N}R.\end{equation}
The problem becomes much more complicated. So we prefer to study
it later.
\section{Kaluza Klein Reduction}
\subsection{Neglect the Dependence of Reduced Dimension}
In nowadays string theorists think that the four dimensional
space-time physics is reduced from a 11-dimensional M-theory
through Kaluza-Klein mechanism. Most naturally the reduced
dimensions are space-like, so we use $\epsilon=1$.\\ When the
compactifying radius is very small, the corresponding massive
fields can be neglected we then assume that the metric is
independent of reduced dimensions, and
\begin{equation}\tilde{\nabla}_\gamma\,h_{\alpha\beta}=0,\qquad\tilde{\nabla}_\gamma\,g_{AB}=0\end{equation}
operator $\tilde{\nabla}_\gamma$ is equivalent to $\nabla_\gamma$,
 Gauss-Codazzi equations become quite simple. In fact
eq.(49)eq.(50) and eq.(52)are changed into following three
equations respectively
\begin{equation}h_C^\zeta\,h_\alpha^Ah_\beta^Bh_\gamma^D{R_{ADB}}^C=\mathbf{{R}_{\alpha\gamma\beta}}^\zeta-
\frac{1}{2N^2}\mathcal{F}_\beta{}^\zeta\mathcal{F}_{\alpha\gamma}+\frac{1}{4N^2}\mathcal{F}_{\gamma\beta}\mathcal{F}_\alpha{}^\zeta
-\frac{1}{4N^2}\mathcal{F}_{\alpha\beta}\mathcal{F}_\gamma{}^\zeta\end{equation}\\
\begin{equation}h^{AB}h^{CD}R_{ADBC}=R-2R_{ab}n^an^b=\mathbf{R}-\frac{3\epsilon}{4N^2}\mathcal{F}_{\alpha\gamma}\mathcal{F}^{\alpha\gamma}\end{equation}\\
\begin{equation}R=\mathbf{R}-2\mathbf{b}(\partial^\alpha\partial_\alpha\phi+\mathbf{b}\partial^\alpha\phi\partial_\alpha\phi)-
\frac{\epsilon}{4N^2}\mathcal{F}_{\alpha\beta}\mathcal{F}^{\alpha\beta}.\end{equation}
And eq.(61) is simplified to
\begin{equation}\tilde{\nabla}_\alpha\,K_\beta{}^\alpha+2\mathbf{b}K_{\alpha\beta}\partial^\alpha\phi=-R_{BA}h_\beta^Bn^A\end{equation}\\
\begin{equation}=-N(R_{\beta\,n}-N_\beta\,R_{nn})\quad
(K_{\alpha\beta}=\frac{-1}{2N}\mathcal{F}_{\alpha\beta},\quad
N_\beta=\mathcal{A}_\beta,N=e^{-\mathbf{b}\phi})\end{equation} By
use of the "vielbein" method or direct calculation the equivalent
form of Gauss equation, the reduction formula, has already
presented in literature (see for example
\cite{Polchinski}\cite{deWit}\cite{Lu} and references therein).
\subsection{Combine With the Fiber Bundle}
Gauss-Codazzi equations can be set up in distinct regions
(exactly, neighborhoods) and there is a transition function
matching them up. As an example let us examine the 11 dimensional
Kaluza-Klein monopole
\begin{equation}ds_{11}^2=h_{\alpha\beta}dx^\alpha\,dx^\beta+\frac{1}{N^2}(dx^{10}+\mathcal{A}^\pm)^2\end{equation}
in which $$N=e^{-\frac{4}{3}\phi}$$ is a functional of dilaton
field $\phi$. Set $x_\alpha=(x_\mu,y_i)$, $\mu=0,\cdots,7;\quad
i=1,2,3$
\begin{equation}\mathcal{A}^\pm=\frac{Q_m}{2r(y_3+r)}(-y_2dy_1+y_1dy_2)=\frac{1}{2}Q_m(\pm1-\cos\theta)d\phi\end{equation}
($r=\sqrt{y_iy_i}$) is the monopole in "south" and "north" regions
(semisphere) respectively \cite{Wu}. Therefore we get in these
respective regions, Gauss and Codazzi equations with
$N_\alpha=\mathcal{A}_\alpha^+$ and  $\mathcal{A}_\alpha^-$. The
transition function is the well-known gauge transformation from
$\mathcal{A}^\mp$ to $\mathcal{A}^\pm$. Of course, when we neglect
the dependence of the metric on reduced dimensions, the difference
in Gauss equation is trivial (because that Gauss equation depends
only on a functional of $\mathcal{A},\mathcal{F}$, but not
on $\mathcal{A}$ itself).\\
By a conformal transformation the 10 dimensional part of eq.(72)
is a D6 brane metric (in string frame)
\begin{equation}ds_{10}^2=N^{-1}h_{\alpha\beta}dx^\alpha\,dx^\beta=N^{-1}\eta_{\mu\nu}dx^\mu\,dx^\nu+Ndy^idy^i,\end{equation}
and
\begin{equation}\mathcal{F}_2=e^{-\frac{3}{2}\phi}*(dx^7\wedge\,dN^{-2})\end{equation}\\
\begin{equation}=\frac{1}{2r^3}\epsilon_{ijk}y^idy^jdy^k=d\mathcal{A}^\pm.\end{equation}
\subsection{Reduced dimensions larger than 1}
Because of the complexity we will continue the assumption that the
metric is independent of reduced dimensions Thus we can easily
prove that equations
$$\tilde{\nabla}_\gamma\,h_{\alpha\beta}=0,\qquad\tilde{\nabla}_\gamma\,g_{AB}=0$$
are still hold in higher reduced dimensions. So we can use
$g_{AB}$ and $h_{\alpha\beta}$ to raise and lower indices of whole
space and subspace respectively. And we may omit tilde symbol on
the covariant derivative operator. Let us start from a metric
\begin{equation}ds^2=g_{AB}dx^Adx^B=h_{\alpha\beta}dx^\alpha\,dx^\beta+N_{ij}(dy^i+N_\alpha^idx^\alpha)(dy^j+N_\beta\,dx^\beta)\end{equation}
\begin{equation}N_{ij}=N_{ji}\end{equation}
in which we have used $i,j,\cdots$ to denote the indices of
reduced dimensions, and they are Euclidean. Then, the metric has
the form
\begin{equation}g_{AB}=h_{AB}+N_{ij}n_A^in_B^j.\qquad\,A=(\alpha,j)\end{equation}
where
\begin{equation}n_A^i=(N_\alpha^i,\delta_j^i),\qquad\,n^{iA}=(0,N^{-1\,ij})\end{equation} and
\begin{equation}n_A^in_j^A=N^{-1}{}_j{}^i,\qquad h_{AB}n_i^B=0=h^{AB}n_B^i,\end{equation}
explicitly we have for example
\begin{equation}h^{AB}=\left(\begin{array}{cc}h^{\alpha\beta}&-N^{j\alpha}\\-N^{i\beta}&N_\alpha^iN^{j\alpha}\end{array}\right),\qquad
\,N_i^jn_j^An_B^i=\left(\begin{array}{cc}0&0\\N_\beta^i&\delta_j^i\end{array}\right).\end{equation}
Along the same line as in the derivation of eq.(27) the Gauss
formula becomes
\begin{equation}\nabla_\alpha\,h_\beta^A=K_{\alpha\beta}{}^in_i^A,\end{equation}
where
\begin{equation}K_{\alpha\beta}{}^i\equiv(\partial_\alpha\,h_\beta^D+\Gamma_{BC}^Dh_\alpha^Bh_\beta^C)n_D^jN_j^i\equiv
\tilde{K}_{\alpha\beta}{}^jN_j^i.\end{equation} Because that
\begin{equation}h_\beta^A\nabla_\alpha\,n_A^i=h_\beta^A(\partial_\alpha\,n_A^i-\Gamma_{BA}^Ch_\alpha^Bn_C^i),\end{equation}
\begin{equation}=-(\partial_\alpha\,h_\beta^A+\Gamma_{BC}^Ah_\alpha^Bh_\beta^C)n_A^i=-K_{\alpha\beta}{}^jn_j^An_A^i
=-\tilde{K}_{\alpha\beta}{}^i,\end{equation} and
\begin{equation}-h_D^\beta\,K_{\alpha\beta}{}^i=(\delta_D^A-N_j^kn_k^An_D^j)\nabla_\alpha\,n_A^i=\nabla_\alpha\,n_D^i-n_D^jL_{\alpha\,j}{}^i,\end{equation}
here we have defined
\begin{equation}{L_{\alpha\,j}}^i\equiv\,N_j^kn_k^A\nabla_\alpha\,n_A^i\equiv\,N_j^k{\tilde{L}_{\alpha\,k}}^i.\end{equation}
Therefore the Weingarten formula can be written as
\begin{equation}\nabla_\alpha\,n_A^i=n_A^jL_{\alpha\,j}{}^i-h_A^\beta\,K_{\alpha\beta}{}^j{N^{-1}{}_j}^i=
-h_A^\beta\tilde{K}_{\alpha\beta}{}^i+N_k^jn_A^k\tilde{L}_{\alpha\,j}{}^i.\end{equation}
Since
\begin{eqnarray*}\lefteqn{(\nabla_\alpha\,n_A^i)n_j^A+n_A^i\nabla_\alpha\,n_j^A=}\\
  & &{-h_A^\beta\tilde{K}_{\alpha\beta}{}^in_j^A+N_k^ln_A^k\tilde{L}_{\alpha\,l}{}^in_j^A-
     n_A^ih^{A\gamma}\tilde{K}_{\alpha\gamma\,j}+n_A^iN_k^l\tilde{L}_\alpha{}^k{}_jn_l^A=}\end{eqnarray*}\begin{eqnarray}
  & &{\tilde{L}_{\alpha\,j}{}^i+\tilde{L}_\alpha{}^i{}_j}.\end{eqnarray}
So the tensor L satisfies the relation
\begin{equation}\tilde{L}_{\alpha\,ji}+\tilde{L}_{\alpha\,ij}=\partial_\alpha\,N^{-1}{}_{ij}.\end{equation}
In the following we will derive the Gauss equation. Since
\begin{equation}\nabla_\alpha\nabla_\beta\,h_\gamma^A-\nabla_\beta\nabla_\alpha\,h_\gamma^A=
h_\alpha^Ch_\beta^Dh_\gamma^BR^A{}_{BCD}-h_\delta^A\mathbf{R}^\delta{}_{\gamma\alpha\beta}+
\Gamma_{BC}^A(\partial_\alpha\,h_\beta^B-\partial_\beta\,h_\alpha^B)h_\gamma^C,\end{equation}
and
\begin{equation}h_A^\delta\nabla_\alpha\nabla_\beta\,h_\gamma^A=-(\nabla_\alpha\,h_A^\delta)(\nabla_\beta\,h_\gamma^A)\end{equation}\\
\begin{equation}=N^{-1}{}^i{}_jK_\alpha{}^\delta{}_iK_{\beta\gamma}{}^j=-N_j^i\tilde{K}_\alpha{}^\delta{}_i\tilde{K}_{\beta\alpha}{}^j,\end{equation}
hence
\begin{equation}h_A^\delta\,h_\alpha^Ch_\beta^Dh_\gamma^BR^A{}_{BCD}=\mathbf{R}^\delta{}_{\gamma\alpha\beta}-
N^{-1}{}^i{}_j(K_\alpha{}^\delta{}_iK_{\beta\gamma}{}^j-K_\beta{}^\delta{}_iK_{\alpha\gamma}{}^j)+(\Gamma_{i\gamma}^\delta-
N_\gamma^j\Gamma_{ij}^\delta)\mathcal{F}_{\alpha\beta}{}^i\end{equation}
in which
\begin{equation}\mathcal{F}_{\alpha\beta}{}^i=\partial_\alpha\,N_\beta^i-\partial_\beta\,N_\alpha^i,\end{equation}
and
\begin{equation}\Gamma_{i\gamma}^\delta-N_\gamma^j\Gamma_{ij}^\delta
=\frac{1}{2}[g^{\delta\alpha}(\partial_\gamma\,g_{i\alpha}-\partial_\alpha\,g_{i\gamma}+N_\gamma^j\partial_\alpha\,g_{ij})+
g^{\delta\,j}\partial_\gamma\,g_{ij}]=\frac{1}{2}N_{ij}h^{\delta\alpha}\mathcal{F}_{\gamma\alpha}{}^j.\end{equation}
At the end we obtain Gauss equation
\begin{equation}h_\alpha^Ah_\beta^Bh_\gamma^Ch_\delta^DR_{ABCD}=\mathbf{R}_{\alpha\beta\gamma\delta}+
N^{-1}{}_j^i(K_{\alpha\delta\,i}K_{\beta\gamma}{}^j-K_{\beta\delta\,i}K_{\alpha\gamma}{}^j)-
\frac{1}{2}N_{ij}\mathcal{F}_{\alpha\beta}{}^i\mathcal{F}_{\gamma\delta}{}^j.\end{equation}\\
To get Codazzi equation and Ricci equation we note that the
following equation is available
\begin{equation}\nabla_\beta\nabla_\alpha\,n_i^A-\nabla_\alpha\nabla_\beta\,n_i^A=R^A{}_{BDC}h_\alpha^Ch_\beta^Dn_i^B+
\Gamma_{BC}^A(\partial_\beta\,h_\alpha^B-\partial_\alpha\,h_\beta^B)n_i^C.\end{equation}
By using of Weingarten formula we know
\begin{equation}\nabla_\beta\nabla_\alpha\,n_i^A=-(\nabla_\beta\tilde{K}_\alpha{}^\gamma{}_i)h_\gamma^A-\tilde{K}_\alpha{}^\gamma{}_iK_{\beta\gamma}{}^jn_j^A
+(\nabla_\beta\,L_\alpha{}^j{}_i)n_j^A+L_\alpha{}^j{}_i(-\tilde{K}_\beta{}^\gamma{}_jh_\gamma^A+L_\beta{}^k{}_jn_k^A),\end{equation}
thus
\begin{eqnarray*}\lefteqn{(\nabla_\beta\nabla_\alpha-\nabla_\alpha\nabla_\beta)n_i^A=}\\
  & &{-(\nabla_\beta\,K_\alpha{}^\gamma{}_i-\nabla_\alpha\tilde{K}_\beta{}^\gamma{}_i)h_\gamma^A-
     (\tilde{K}_\alpha{}^\gamma{}_iK_{\beta\gamma}{}^j-\tilde{K}_\beta{}^\gamma{}_iK_{\alpha\gamma}{}^j)n_j^A}\\
  & &{+(\nabla_\beta\,L_\alpha{}^j{}_i-\nabla_\alpha\,L_\beta{}^j{}_i)n_j^A-L_\alpha{}^j{}_i\tilde{K}_\beta{}^\gamma{}_j-
     L_\beta{}^j{}_i\tilde{K}_\alpha{}^\gamma{}_j)h_\gamma^A+}\end{eqnarray*}\begin{eqnarray}
  & &{(L_\alpha{}^j{}_iL_\beta{}^k{}_j-L_\beta{}^j{}_iL_\alpha{}^k{}_j)n_k^A}\end{eqnarray}\\
\begin{equation}=R^A{}_{BCD}h_\beta^Ch_\alpha^Dn_i^B+\Gamma_{BC}^A(\partial_\beta\,h_\alpha^C-\partial_\alpha\,h_\beta^C)n_i^B.\end{equation}
Note that
\begin{equation}\Gamma_{jk}^\gamma=-\frac{1}{2}\partial^\gamma\,g_{jk},\qquad
\Gamma_{kj}^l=\frac{1}{2}N_\gamma^l\partial^\gamma\,g_{jk},\end{equation}
at last we obtain Codazzi equation
\begin{equation}\nabla_\alpha\tilde{K}_{\beta\gamma\,i}-\nabla_\beta\tilde{K}_{\alpha\gamma\,i}-
L_\alpha{}^j{}_i\tilde{K}_{\beta\gamma\,j}+L_\beta{}^j{}_i\tilde{K}_{\alpha\gamma\,j}\\
=-R_{ABCD}h_\alpha^Ch_\beta^Dn_i^B-\frac{1}{2}(\partial_\gamma\,N_{jk})N^{-1}{}_i^k\mathcal{F}_{\alpha\beta}^j,\end{equation}
and Ricci equation
\begin{eqnarray*}\lefteqn{(\nabla_\beta\,L_\alpha{}^j{}_i-\nabla_\alpha\,L_\beta{}^j{}_i)N^{-1}{}_j^l+
     (L_\alpha{}^j{}_iL_\beta{}^k{}_j-L_\beta{}^j{}_iL_\alpha{}^k{}_j)N^{-1}{}_k^l-}\\
  & &{(\tilde{K}_\alpha{}^\gamma{}_iK_{\beta\gamma}{}^j
     -\tilde{K}_\beta{}^\gamma{}_iK_{\alpha\gamma}{}^j)N^{-1}{}_j^l=}\end{eqnarray*}\begin{eqnarray}
  & &{R^A{}_{BCD}h_\beta^Ch_\alpha^Dn_i^Bn_A^l-\frac{1}{2}(\delta_\delta^\gamma-
     N_\delta^l)(\partial^\delta\,N_{jk})N^{-1}{}_i^k\mathcal{F}_{\alpha\beta}{}^j}.
\end{eqnarray}

\section{Discussion}
In this paper we have derived variations of Gauss-Codazzi-Ricci
equations, but there are several questions we need to research
further.Firstly, we want to know what geometric meaning the tensor
$K_{\alpha\beta}$ has? Ordinary, the corresponding tensor
presented in Gauss-Codazzi-Ricci equations is the extrinsic
curvature, which is connected with the second fundamental form. In
three dimensional Euclidean space half of second fundamental form
is the principle part of the departure from tangent plane to a
point which is in the neighborhood of the tangent point on the
surface .
\\Next, as mentioned in sec.5 we have to perform the Hamiltonian
formulation of general relativity in new metric construction.\\
Thirdly, if metric depends on reduced dimensions in some special
way, such as the spherical reductions \cite{Duff} it must be very
interesting to know what new feature will arise.


\appendix{Christoffel Symbols for One Reduced Dimension}
\begin{appendix}
\begin{equation}\Gamma_{\alpha\beta}^\gamma=\mathbf{\Gamma}_{\alpha\beta}^\gamma+
\frac{\epsilon\,h^{\gamma\delta}}{2N^2}(N_\alpha\mathcal{F}_{\beta\delta}+N_\beta\mathcal{F}_{\alpha\delta}-
2\mathbf{b}N_\alpha\,N_\beta\partial_\delta\phi)+\frac{N^\gamma}{2N^2}(\mathcal{H}_{\alpha\beta}+2\epsilon\mathbf{b}N_\alpha\,N_\beta\partial_n\phi)\end{equation}\\
\begin{eqnarray*}\lefteqn{\Gamma_{\alpha\beta}^n=}\\
  & &{-N_\gamma\mathbf{\Gamma}_{\alpha\beta}^\gamma-\frac{\epsilon\,N^\gamma}{2N^2}(N_\alpha\mathcal{F}_{\beta\gamma}
     +N_\beta\mathcal{F}_{\alpha\gamma}-2\mathbf{b}N_\alpha\,N_\beta\partial_\gamma\phi)+1/2N^2[\partial_\alpha(N_\beta/N^2)}\end{eqnarray*}\begin{eqnarray}
  & &{+\partial_\beta(N_\alpha/N^2)]-1/{2N^2}(N_\gamma\,N^\gamma+\epsilon\,N^2)(\mathcal{H}_{\alpha\beta}
     +2\epsilon\mathbf{b}N_\alpha\,N_\beta\partial_n\phi)}\end{eqnarray}\\
\begin{equation}\Gamma_{n\alpha}^\beta=\frac{h^{\beta\gamma}}{2N^2}[\epsilon(\mathcal{F}_{\alpha\gamma}-
2\mathbf{b}N_\alpha\partial_\gamma\phi)+\mathcal{H}_{\gamma\alpha}+2\epsilon\mathbf{b}N_\gamma\,N_\alpha\partial_n\phi]\end{equation}\\
\begin{equation}\Gamma_{n\alpha}^n=-\frac{\epsilon\,N^\gamma}{2N^2}(\mathcal{F}_{\alpha\gamma}-
2\mathbf{B}N_\alpha\partial_\gamma\phi)+\mathbf{b}\partial_\alpha\phi-\frac{N^\gamma}{2N^2}(\mathcal{H}_{\gamma\alpha}+
2\epsilon\mathbf{b}N_\gamma\,N_\alpha\partial_n\phi)\end{equation}\\
\begin{equation}\Gamma_{nn}^\gamma=\frac{\epsilon}{N^2}[-\mathbf{b}(\partial^\gamma\phi-N^\gamma\partial_n\phi)+h^{\gamma\alpha}\partial_nN_\alpha]\end{equation}\\
\begin{equation}\Gamma_{nn}^n=\frac{\epsilon}{N^2}[\mathbf{b}(N_\gamma\partial^\gamma\phi-(N_\gamma\,N^\gamma-
\epsilon\,N^2)\partial_n\phi)-N^\gamma\partial_nN_\gamma]\end{equation}
\end{appendix}.
\end{document}